\documentclass[runningheads]{llncs}
\usepackage{graphicx}
\usepackage{multirow}
\usepackage{xcolor}
\usepackage{caption}
\usepackage{placeins}

\usepackage[ruled,vlined]{algorithm2e}
\begin{document}
\title{Benchmarking 
Data Lakes Featuring 
Structured and Unstructured 
Data 
with DLBench
}
\titlerunning{Benchmarking Data Lakes with DLBench}
%
\author{Pegdwendé N. Sawadogo\inst{1}
\and
Jérôme Darmont\inst{1}
}
\authorrunning{P.N. Sawadogo et J. Darmont}
%
\institute{Université de Lyon, Lyon 2, UR ERIC \\
5 avenue Pierre Mendès France, F69676 Bron Cedex, France
\email{\{pegdwende.sawadogo,jerome.darmont\}@univ-lyon2.fr}\\
}
\maketitle              
\begin{abstract}
In the last few years, the concept of data lake has become trendy for data storage and analysis. Thus, several approaches have been proposed to build data lake systems. However, these proposals are difficult to evaluate as there are no commonly shared criteria for comparing data lake systems. Thus, we introduce DLBench, a benchmark to evaluate and compare data lake implementations that support textual and/or tabular contents. More concretely, we propose a data model made of both textual and CSV documents, a workload model composed of a set of various tasks, as well as a set of performance-based metrics, all relevant to the context of data lakes. As a proof of concept, we use DLBench to evaluate an open source data lake system we previously developed.

\end{abstract}
\section{Introduction}

Over the last decade, the concept of data lake has emerged as a reference for data storage and exploitation. A data lake is  a large repository for storing and analyzing data of any type and size, kept in their raw format~\cite{Dixon2010}. Data access and analyses from data lakes largely rely on metadata~\cite{Maccioni2018}, making data lakes flexible enough to support a broader range of analyses than traditional data warehouses. Data lakes are thus handy for both data retrieval and data content analysis. 

However, the concept of data lake  still 
lacks standards~\cite{Russom2017}. Thus, there is no commonly shared approach to build, nor to evaluate a data lake. Moreover, existing data lake architectures are often evaluated in diverse and specific ways, and are hardly comparable with each other. Therefore, there is a need of benchmarks to allow objective and comparative evaluation of data lake implementations. 
There are several benchmarks for big data systems in the literature, but none of them considers  the wide range of possible analyses in data lakes. 

Hence, we propose in this paper the Data Lake Benchmark (\textsc{DLBench}) to evaluate data management performance in data lake systems. We particularly focus in this first instance on textual and tabular contents, which are often included in data lakes.  \textsc{DLBench} is data-centric, i.e., it focuses on a data management objective, regardless of the underlying technologies~\cite{Darmont2019}. We also designed it with Gray's criteria for a ``good" benchmark in mind, namely  relevance, portability, simplicity and scalability~\cite{Gray1993}.

More concretely, 
 \textsc{DLBench} features a data model that generates textual and tabular documents. By tabular documents, we mean spreadsheet or Comma Separated Value (CSV) files whose integration and querying is a common issue in data lakes. A scale factor parameter $SF$ allows to vary data size in predetermined proportions. 
\textsc{DLBench} also features a workload model, i.e., a set of analytical operations relevant to the context of data lakes with textual and/or tabular content.  
Finally, we propose a set of performance-based metrics to evaluate such data lake implementations, as well as  an execution protocol to execute the workload model on the data model and compute the metrics. 

The remainder of this paper is organized as follows. In Section~\ref{sec:related.works}, we show how \textsc{DLBench} differs from existing benchmarks. 
In Section~\ref{sec:description}, we provide \textsc{DLBench}'s full specifications. 
In Section~\ref{sec:poc}, we exemplify how \textsc{DLBench} works and the insights it provides.  Finally, in Section~\ref{sec:conclusion}, we conclude this paper and present research perspectives.

\section{Related Works}
\label{sec:related.works}

Benchmarking data lakes mainly relates to two benchmark categories, namely big data and text benchmarks. In this section, we present recent works in these categories and discuss their limitations with respect to our benchmarking objectives. 

\subsection{Big Data Benchmarks}
Big data systems are so diverse that each of big data benchmarks in the literature  only target a part of big data  requirements~\cite{Bajaber2020}.
The \textit{de facto} standard TPC-H~\cite{TPCH} and TPC-DS~\cite{TPCDS} issued by the Transaction Processing Performance Council are still widely used to benchmark traditional business intelligence systems. 
They provide data models that reflect a typical data warehouse, as well as a set of typical business queries, mostly in SQL. 
\textsc{BigBench}~\cite{Ghazal2013} is another reference benchmark that addresses SQL querying on data warehouses. In addition, \textsc{BigBench} adaptations~\cite{Ghazal2017,Ivanov2020} include more complex big data analysis tasks, namely sentiment analysis over short texts. 

TPCx-HS~\cite{TPCHS} is a quite different benchmark that aims to evaluate systems running on Apache Hadoop\footnote{\url{https://hadoop.apache.org/}} or Spark\footnote{\url{http://spark.apache.org/}}. For this purpose, 
only a sorting workload helps measuring performances. \textsc{HiBench}~\cite{Huang2010} also evaluates Hadoop/Spark systems, but with a broader range of workloads, i.e., ten workloads including SQL aggregations and joins, classification, clustering and sorts~\cite{Ivanov2015}. In the same line, TPCx-AI~\cite{TPCAI}, which is still in development, includes more  analysis tasks relevant to big data systems, such as advanced machine learning tasks for fraud detection and product rating. 


\subsection{Textual Benchmarks}
In this category, we consider big data benchmarks with a consequent part of text on  one hand, and purely textual benchmarks on the other hand. 
\textsc{BigDataBench}~\cite{Wang2014} is a good representative of the first category. It indeed includes a textual dataset made of Wikipedia\footnote{https://en.wikipedia.org/} pages, as well as classical information retrieval workloads such as \textit{Sort}, \textit{Grep} and \textit{Wordcount} operations.

One of the latest purely textual benchmarks is \textsc{TextBenDS}~\cite{Truica2021}, which aims to evaluate performances of text analysis and processing systems. For this purpose, \textsc{TextBenDS} proposes a tweet-based data model and two types of workloads, namely \textit{Top-K keywords} and \textit{Top-K documents} operations. Other purely textual benchmarks focus on language analysis tasks, e.g.,  Chinese \cite{Zhu2019} and Portuguese  \cite{Fialho2020}  text recognition, respectively. 

\subsection{Discussion}
None of the aforementioned benchmarks proposes a workload sufficiently extensive to reflect all relevant operations in data lakes. In the case of structured data, most benchmark workloads only consider SQL operations (TPC-H, TPC-DS, \textsc{HiBench}). More sophisticated machine learning operations remain marginal, while they are common analyses in data lakes.  Moreover, the task of finding related data (e.g., joinable tables) is purely missing, while it is a key feature of data lakes.

Existing textual workloads are also insufficient. Admittedly, \textsc{BigDataBench}'s   \textit{Grep} and \textsc{TextBenDS}'s \textit{Top-K documents} operation are relevant for data search. Similarly, \textit{Top-K keywords} and \textit{WordCount} are relevant to assess documents aggregation~\cite{Huang2010,Truica2021}. However, other operations such as finding most similar documents or clustering documents should also be considered. 

Thus, our \textsc{DLBench} benchmark  stands out, with a broader workload that features both data retrieval and data content analysis operations. 
\textsc{DLBench}'s data model also differs from most big data benchmarks as it provides  raw tabular files, inducing an additional data integration challenge. Moreover, \textsc{DLBench} includes a set of long textual documents that induces a different challenge than short texts such as  tweets~\cite{Truica2021} and Wikipedia articles~\cite{Wang2014}.
 Finally, \textsc{DLBench} is data-centric, unlike big data benchmarks that focus on a particular technology, e.g.,  \textsc{TPCx-HS} and \textsc{HiBench}.



\section{Specification of \textsc{DLBench}}
\label{sec:description}

In this section, we first provide a thorough description of \textsc{DLBench}’s data 
and workload model. 
Then, we propose a set of metrics 
and introduce an assessment protocol to evaluate and/or compare systems using \textsc{DLBench}. 

\subsection{Data Model}
\label{subsec:bench.data}

\subsubsection{Data Description}
\textsc{DLBench} includes two types of data to simulate a data lake: textual documents and tabular data. 
Textual documents are scientific articles that span from few to tens of pages. 
They are written in French and English and their number amounts to 50,000. Their overall volume is about 62~GB. 

Tabular data are synthetically derived from a few CSV files containing Canadian government open data.  
Although such data are often considered as structured, they still need integration to be queried and analyzed effectively. \textsc{DLBench} features up to 5,000 tabular files 
amounting to about 1,4 GB of data. 

The amount of data in the benchmark can be customised through scale factor parameter $SF$, 
which is particularly useful to measure a system's performance when data volume increases. 
$SF$ ranges from 1 to 5. 
Table~\ref{tab:scale.factors} describes the actual amount of data obtained with values of $SF$.  
\begin{table*}[hbt]
        \centering
        \caption{Amount of data per $SF$ value}
        \label{tab:scale.factors}
        \begin{tabular}{|l|r|r|r|r|r|}
            \hline
            Scale factors & $SF=1$ & $SF=2$  & $SF=3$ & $SF=4$  & $SF=5$\\
            \hline \hline
            Nb. of textual documents & 10,000 & 20,000 & 30,000 & 40,000 & 50,000  
            \\ \hline
           Nb. of tabular files & 1,000 & 2,000 & 3,000 & 4,000 & 5,000
            \\ \hline
           Textual documents' size (GB) & 8.0 & 24.9 & 37.2 & 49.6 & 62.7
            \\ \hline
           Tabular files' size (GB) & 0.3 & 0.6 & 0.8 & 1.1 & 1.4  
            \\
            \hline
        \end{tabular}
    \end{table*}
    
\textsc{DLBench}'s data come with metadata catalogues that can serve in data integration. More concretely, we generate from textual documents catalogue information on \textit{year}, \textit{language} and \textit{domain} (discipline) to which each document belongs. Similarly, we associate in the tabular file catalogue a \textit{year} with each file. 
This way, we can separately query each type of data  through its specific metadata. We can also jointly query textual documents and tabular files through the \textit{year} field.


Eventually, textual documents are generated 
independently from tabular files. Therefore, each type of data can be used apart from the other. In other words, \textsc{DLBench} can be used to assess a system that contains either textual documents only, tabular files only, or both. When not using both types of data, the workload model must be limited to its relevant part.


\subsubsection{Data Extraction}
We extract textual data from HAL\footnote{\url{https://hal.archives-ouvertes.fr/}}, a French open data repository dedicated to scientific document diffusion. We opted for scientific documents as most are long  enough to provide complexity and reflect most actual use cases in textual data integration systems, in contrast with shorter documents such as reviews and tweets. 

Although HAL's access is open, we are not allowed to redistribute 
data extracted from HAL. Thus, we provide instead a script that extracts a user-defined amount of documents. This script and a usage guide are available online for reuse\footnote{\url{https://github.com/Pegdwende44/DLBench}}. 
Amongst all available documents in HAL, we restrict to scientific articles whose length is 
homogeneous, which amounts to 50,000 documents. 
While extracting documents, the script also generates the metadata catalogue described above. 

Tabular data are reused from an existing benchmark~\cite{Nargesian2018B}. These are actually a set of 5,000 synthetic tabular data files 
generated from an open dataset stored inside a SQLite\footnote{\url{https://www.sqlite.org/}} database. 
Many of the columns in the tables contain similar data and can therefore be linked, making this dataset suitable to assess structured data integration as performed in data lakes.

We apply on this original dataset\footnote{https://storage.googleapis.com/table-union-benchmark/large/benchmark.sqlite} a script to extract all (or a part) of the tables in the form of raw CSV files. As for textual documents, this second script also generates a metadata catalogue. The script as well as guidelines are available online$^7$. 

\subsection{Workload Model}
\label{subsec:bench.quer}
To assess and compare data lakes across different implementations and systems, some relevant tasks are needed. Thus we specify in this section instances of probable tasks in textual and tabular data integration systems. Furthermore, we translate each task  into concrete, executable queries (Table~\ref{tab:query.instances}).

\subsubsection{Data Retrieval Tasks}
are operations that find data bearing given characteristics. Three main ways are usually exploited to retrieve data in a lake. They are relevant for both tabular data and textual documents. However, we mainly focus on data retrieval from textual documents, as they represent the largest amount of data. 
\begin{enumerate}
\item  \textit{Category filters}
    consist in filtering data using tags or data properties from the metadata catalogue. In other words, it can be viewed as a navigation task. 
    
     \item \textit{Term-based search}
      pertains to find data, with the help of an index, from all data files that contain a set of keywords. Keyword search is especially relevant for textual documents, but may also serve to retrieve tabular data. %
      
     \item \textit{Related data search}
      aims to,  from a specified data file, retrieve similar data. It can be based on, e.g., column similarities in tabular data, or semantic similarity between textual documents.

    \end{enumerate}

\subsubsection{Textual Document Analysis/Aggregation}

tasks work on 
data contents. Although textual documents and tabular data can be explored with the same methods, they require more specific techniques to be jointly analyzed or aggregated in data lakes. 

\begin{enumerate}
\setcounter{enumi}{3}
\item \textit{Document scoring} 
 is a classical information retrieval task that consists in providing a score for each document with respect to how it matches a set of terms. Such scores can be calculated by diverse ways, e.g., with the ElasticSearch~\cite{Elasticsearch2019} scoring algorithm. In all cases, scores depend on the appearance frequency of query terms in the document to score and in the corpus, and also the document's length. This operation is actually very similar to computing top-$k$ documents. 

\item \textit{Document highlights} 
extract a concordance from a corpus. A concordance is a list of snippets 
where a set of terms appear. It is also a classical information retrieval task that provides a sort of summary of documents. 

\item \textit{Document top keywords} 
are another classical way to summarize and aggregate documents~\cite{Ravat2008}. Computing top keywords is thus a suitable task to assess systems handling textual documents. 

\item \textit{Document text mining.} 
In most data lake systems, data are organized in collections,  using tags for example. Here, we propose a data mining task that consists either in  representing each collection of documents with respect to the others, or in grouping together similar collections with respect to their intrinsic vocabularies. In the first case, we propose a Principal Component Analysis (PCA)~\cite{Wold1987} where statistical individuals are document collections. PCA could, for example, out put an average bag of words for each collection. In the second case, we propose a KMeans~\cite{Steinley2006} clustering to detect groups of similar collections. 

\end{enumerate}

\subsubsection{Tabular Data Analysis/Queries}
Finally, we propose specific tasks suitable for integrated tabular files.

\begin{enumerate}
  \setcounter{enumi}{7}
\item \textit{Simple table queries.}  
We first propose to evaluate a data lake system's capacity to answer simple table queries through a query language such as SQL. %
As we are in a context of raw tabular data, language-based querying is indeed an important challenge to address. 

\item \textit{Complex table queries.}  
In line with the previous task, we propose to measure how the system supports advanced queries, namely join and grouping queries. 

\item \textit{Tuple mining.} 
An interesting way to analyze tabular data is either to represent each row with respect to the others or to group together very similar rows. We essentially propose here the same operation as Task \#7 above, except that statistical individuals are table rows instead of textual documents. 
To achieve such an analysis, we only consider numeric values. 

\end{enumerate}

\begin{table}[hbt]
\caption{Query instances}
\begin{center}
\begin{tabular}{|c|l l|}
\hline
    \textbf{Task} & \multicolumn{2}{|l|}{\textbf{Query}}\\
     \hline
     \hline
   \multicolumn{3}{|c|}{\textbf{\textit{Data retrieval}}}\\
   \hline
    \multirow{3}{*}{\#1} & Q1a & Retrieve documents written in \textit{French} \\
            & Q1b & Retrieve documents written in \textit{English} and edited in \textit{December} \\
           & Q1c  & Retrieve documents whose domains are \textit{math} or \textit{info}, 
             written in \textit{English} \\&& and edited in \textit{2010}, \textit{2012} or \textit{2014} \\
    \hline
     \multirow{2}{*}{\#2} & Q2a & Retrieve data files (documents or tables) containing the term \textit{university} \\
        & Q2b & Retrieve data files containing the terms \textit{university}, \textit{science} or \textit{research}\\
     \hline
     \multirow{2}{*}{\#3} & Q3a & Retrieve the top 5 documents similar to any given document \\
         & Q3b & Retrieve 5 tables joinable to table $t\_dc9442ed0b52d69c\_\_\_\_c11\_1\_\_\_\_1$\\ 
     \hline
    \hline
   \multicolumn{3}{|c|}{\textbf{\textit{ Textual Document Analysis/Aggregation}}}\\
     \hline
   \multirow{2}{*}{\#4} & Q4a & Calculate documents scores w.r.t. the terms \textit{university} and \textit{science}   \\
        & Q4b & Calculate documents scores w.r.t. the terms \textit{university}, \textit{research}, \\
        & & \textit{new} and \textit{solution}\\
    \hline
   \multirow{2}{*}{\#5} & Q5a & Retrieve documents concordance w.r.t. the terms \textit{university} and \textit{science}   \\
        & Q5b & Retrieve documents concordance w.r.t. the terms \textit{university}, \textit{science} \\
        & & \textit{new} and \textit{solution} \\
     \hline
      \multirow{1}{*}{\#6} & Q6a  & Find top 10 keywords from all documents (stopwords excluded) \\
     \hline
     \multirow{2}{*}{\#7} & Q7a & Run a PCA with documents merged by \textit{domains}   \\
        & Q7b & Run a 3-cluster KMeans clustering with documents merged by \textit{domains} \\
     \hline
     \hline
   \multicolumn{3}{|c|}{\textbf{\textit{Tabular Data Analysis/Queries}}}\\
     \hline
     \multirow{2}{*}{\#8} & Q8a & Retrieve all tuples from table $t\_e9efd5cda78af711\_\_\_\_c11\_1\_\_\_\_1$    \\
        & Q8b & Retrieve tuples from table $t\_e9efd5cda78af711\_\_\_\_c11\_1\_\_\_\_1$ \\
        & & whose column \textit{PROVINCE} bears the value \textit{BC} \\
     \hline
     \multirow{2}{*}{\#9} & Q9a & Calculate the average of columns \textit{Unnamed: 12}, \textit{13}, and \textit{20} \\  & & from table  $t\_356fc1eaad97f93b\_\_\_\_c15\_1\_\_\_\_1$ grouped by \textit{Unnamed: 2}   \\
        & Q9b & Run a left join query between tables $PED\_SK\_DTL\_SNF\_\_\_\_c7\_0\_\_\_\_1$ \\
        & & and $t\_285b3bcd52ec0c86\_\_\_\_c13\_1\_\_\_\_1$ w.r.t. columns named \textit{SOILTYPE} \\
     \hline
     \multirow{2}{*}{\#10} & Q10a & Run a PCA on the result of query \textit{Q9a}   \\
        & Q10b & Run a 3-cluster KMeans clustering on the result of query \textit{Q9a} \\
     \hline
\end{tabular}
\end{center}
\label{tab:query.instances}
\end{table}

\subsection{Performance Metrics}
\label{subsec:bench.metrics}

In this section, we propose a set of three metrics to compare and assess data lake implementations. 
\begin{enumerate}

\item \textbf{Query execution time}
 aims to measure the time necessary 
 to run each of the 20 query instances from Table~\ref{tab:query.instances} on the tested data lake architecture. This metric actually reports how efficient the lake's metadata system is, as it serves to integrate raw data, and thus make analyses easier and faster. 
In the case where certain queries are not supported, measures are only computed on the supported tasks.

\item \textbf{Metadata size} 
 measures the amount of metadata generated by the system. It allows to balance the execution time with the resulting storage cost.

\item \textbf{Metadata generation time} 
encompasses the generation of all the lake's metadata.  This also serves to balance query execution time.
\end{enumerate}

We did not include other possible metrics such as actually used RAM and CPU because they are hard to measure. However, we recommend interpreting benchmark results while taking into account available RAM and CPU.


\subsection{Assessment Protocol}
\label{subsec:bench.proto}

The three metrics from Section~\ref{subsec:bench.metrics} are measured through an iterative process for each scale factor $SF \in \{1,2,3,4,5\}$. 
Each iteration consists in four steps (Algorithm~\ref{algo.proto}).
\begin{enumerate}
    \item \textbf{Data generation} is achieved with the scripts specified in Section~\ref{subsec:bench.data}$^7$.
    \item \textbf{Data integration.} Raw data now need to be integrated in the data lake system through the generation and organization of metadata. This step is specific to each system, as there are plethora of ways to integrate data in a lake.
    \item \textbf{Metadata size and generation time computing} consists in measuring metrics the total size of generated metadata and the time taken to generate all metadata, with respect to the current $SF$. 
    \item \textbf{Query execution time computation} involves computing the running time of each individual query. To mitigate any perturbation, we average the time of 10 runs for each query instance. 
    Let us notice that all timed executions must be warm runs, i.e., each of the 20 query instances must first be executed once (a cold run not taken into account in the results). 
\end{enumerate}


\begin{algorithm}[hbt]
\SetAlgoLined
\KwResult{metric\_1, metric\_2, metric\_3}
metric\_1  ←  [ ][ ];
 metric\_2  ←  [ ]; 
 metric\_3  ←  [ ]\;
 \For { SF ← 1  {to}  5 } {
    generate\_benchmark\_data(SF)\;
    generate\_and\_organize\_metadata(SF)\;
    metric\_2[SF]  ←  retrieve\_metadata\_generation\_time(SF)\;
    metric\_3[SF]  ←  retrieve\_metadata\_size(SF)\;
    \For { i ← 1  {to}  20 } {
        run\_query(i, SF)\;
        response\_times  ←  [ ]\;
        \For { j ← 1  {to}  10 } {
        
         response\_times[j]  ←  run\_query(i, SF)\;
        }
        metric\_1[SF][i]  ←  average(response\_times)\;
    }
 }
 \caption{Assessment protocol}
 \label{algo.proto}
\end{algorithm}


\section{Proof of Concept}
\label{sec:poc}

\subsection{Overview of AUDAL}
To demonstrate the use of \textsc{DLBench}, we evaluate AUDAL~\cite{Scholly2021}, a data lake system designed as part of a management science project, to allow automatic and advanced analyses on various textual documents (annual reports, press releases, websites, social media posts...) and spreadsheet files (information about companies, stock market quotations...).
The AUDAL system uses an extensive metadata system stored inside MongoDB\footnote{\url{https://www.mongodb.com}}, Neo4J\footnote{\url{https://neo4j.com}}, SQLite$^8$ and ElasticSearch\footnote{\url{https://www.elastic.co}} to support numerous analyses.

AUDAL provides ready-to-use analyses via a representational state transfer application programming interface (REST API) dedicated to data scientists, and also through a Web-based analysis platform designed for business users.

\subsection{Setup and Results}
AUDAL is implemented on a cluster of three VMware virtual machines (VMs). The first VM has a 7-core Intel-Xeon 2.20 GHz processor and 24 GB of RAM. It  runs  the  API and also supports metadata extraction.  
Both  other  VMs  have  a  mono-core  Intel-Xeon 2.20 GHz processor and 24 GB of RAM. 
Each of the three VMs hosts a Neo4J instance, an ElasticSearch instance and a MongoDB instance to store AUDAL’s metadata.

\begin{figure*}[h]
             \begin{minipage}{.33\textwidth}
                \centering
                \includegraphics[width=\linewidth]{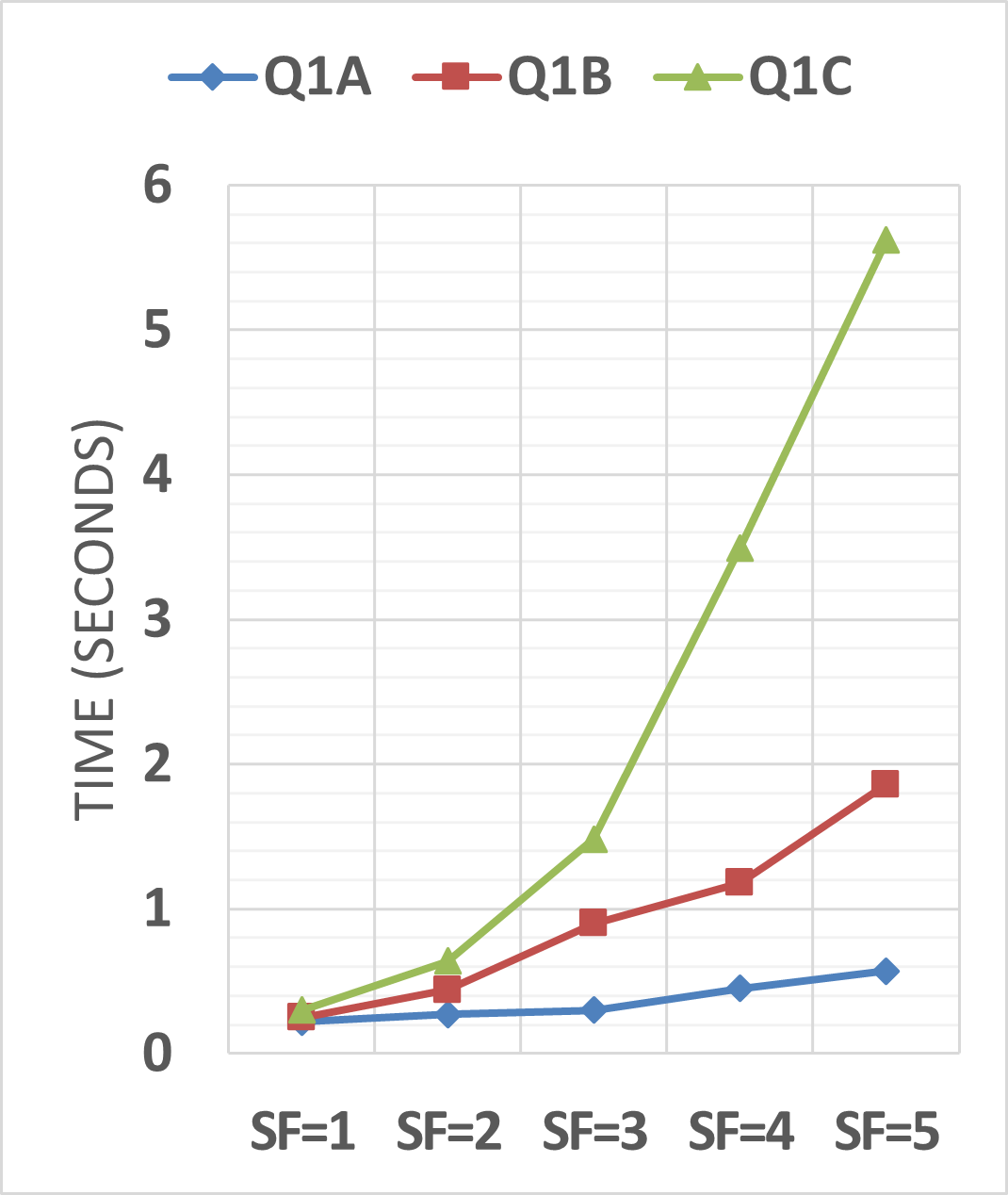}
                \captionsetup{justification=centering}
                \captionof{figure}{Task \#1 \newline response times}
                \label{fig:Q1}
            \end{minipage}%
            \begin{minipage}{.33\textwidth}
                \centering
                \includegraphics[width=\textwidth]{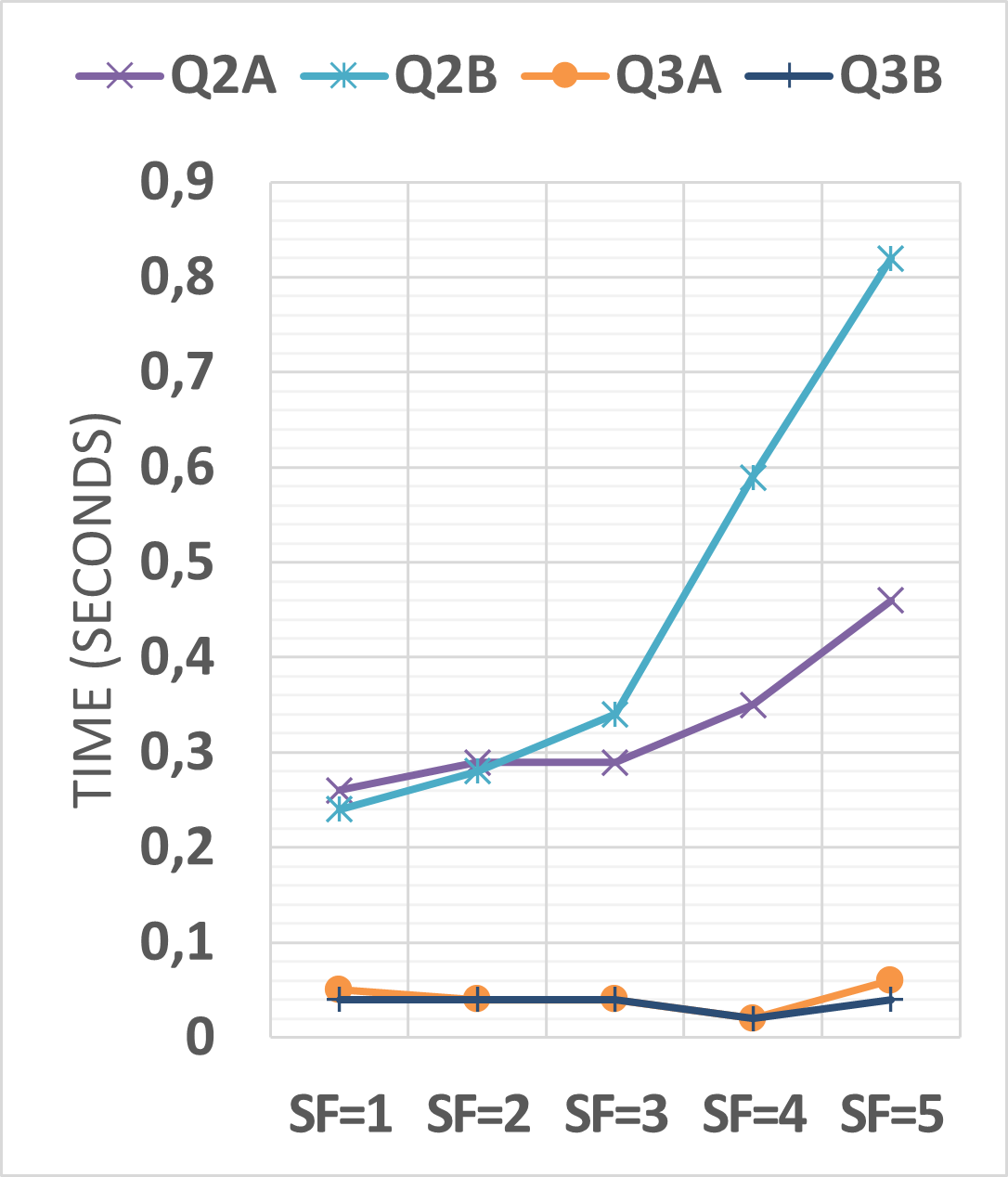} 
                \captionsetup{justification=centering}
                \captionof{figure}{Tasks \#2 \& \#3 \newline response times}
                \label{fig:inter}
            \end{minipage}
            \begin{minipage}{.33\textwidth}
                \centering
                \includegraphics[width=\textwidth]{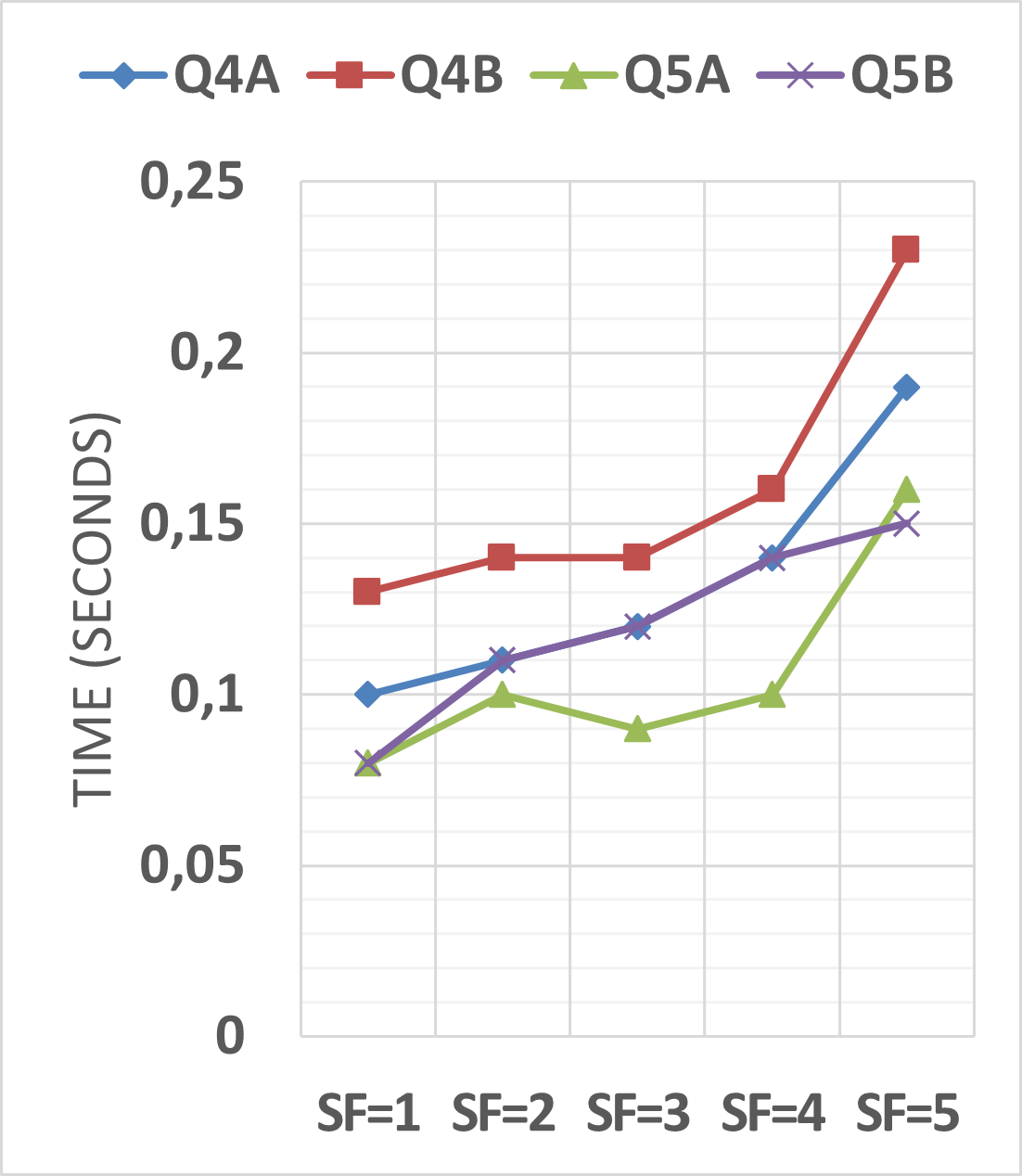} 
                \captionsetup{justification=centering}
                \captionof{figure}{Tasks \#4 \& \#5 \newline response times}
                \label{fig:inter}
            \end{minipage}
        
            \begin{minipage}{.33\textwidth}
                \centering
                \includegraphics[width=\linewidth]{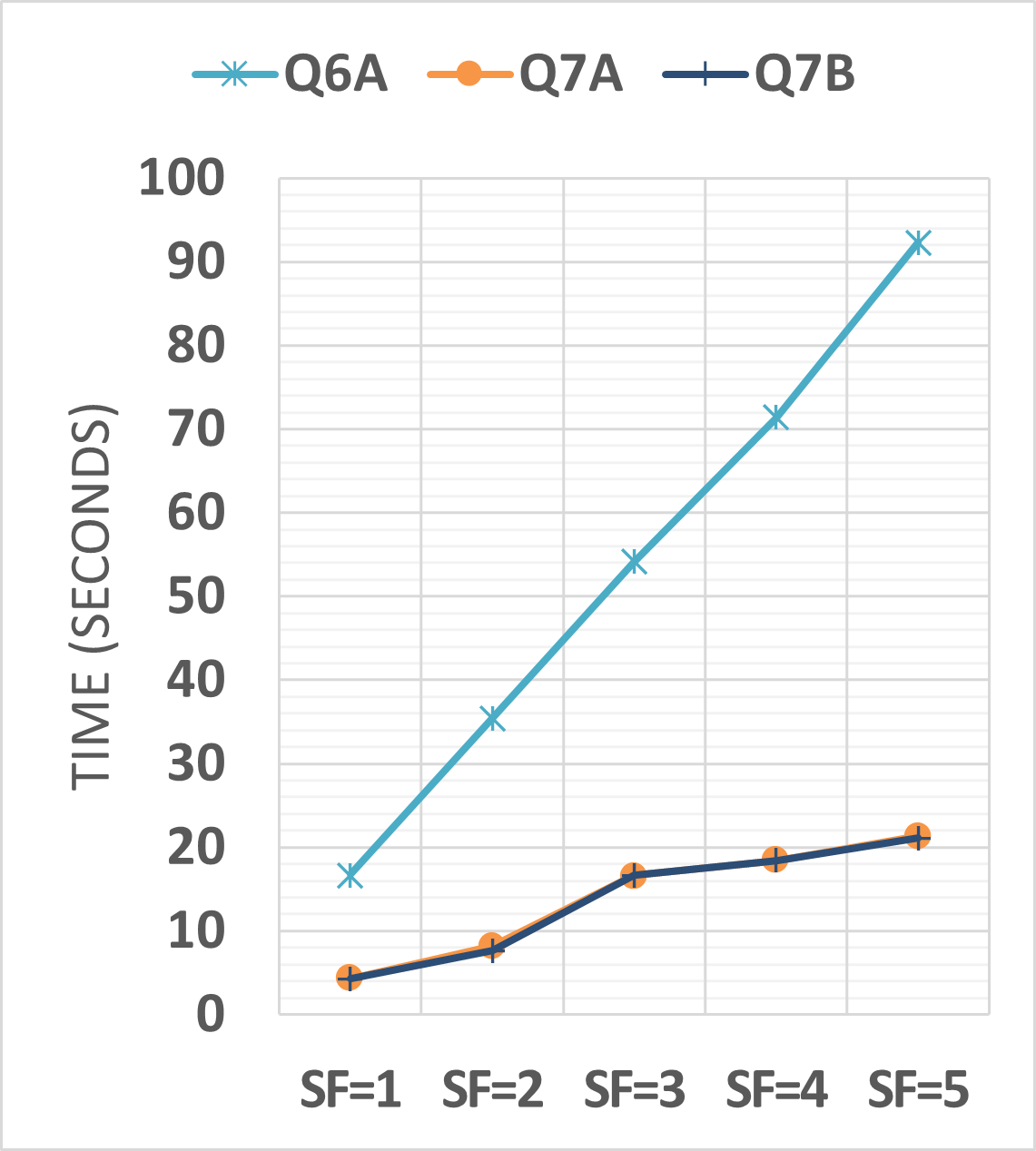}
                \captionsetup{justification=centering}
                \captionof{figure}{Tasks \#6 \& \#7 \newline response times}
                \label{fig:intra}
            \end{minipage}%
            \begin{minipage}{.33\textwidth}
                \centering
                \includegraphics[width=\textwidth]{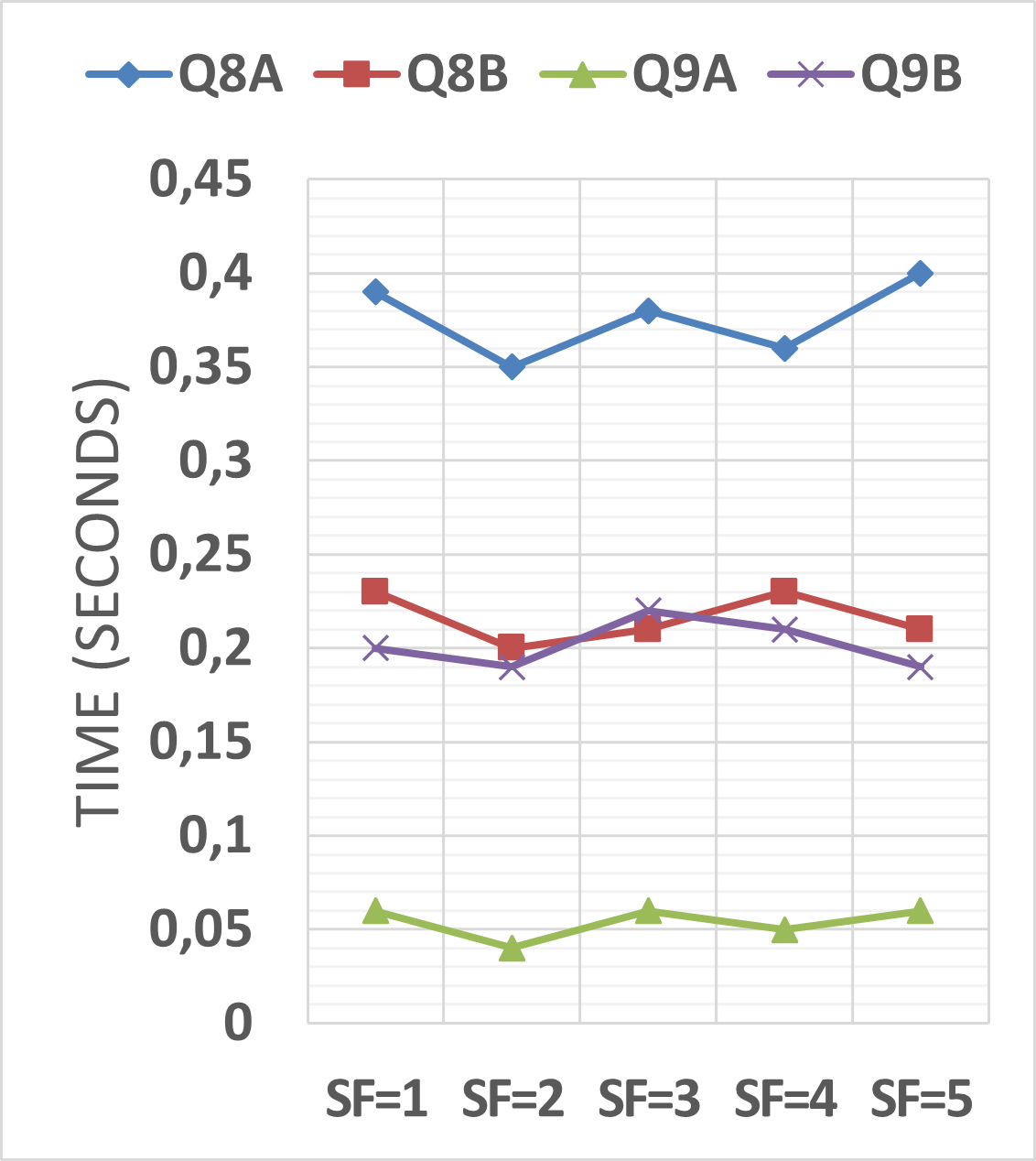} 
                \captionsetup{justification=centering}
                \captionof{figure}{Tasks \#8 \& \#9 \newline response times}
                \label{fig:inter}
            \end{minipage}
            \begin{minipage}{.33\textwidth}
                \centering
                \includegraphics[width=\textwidth]{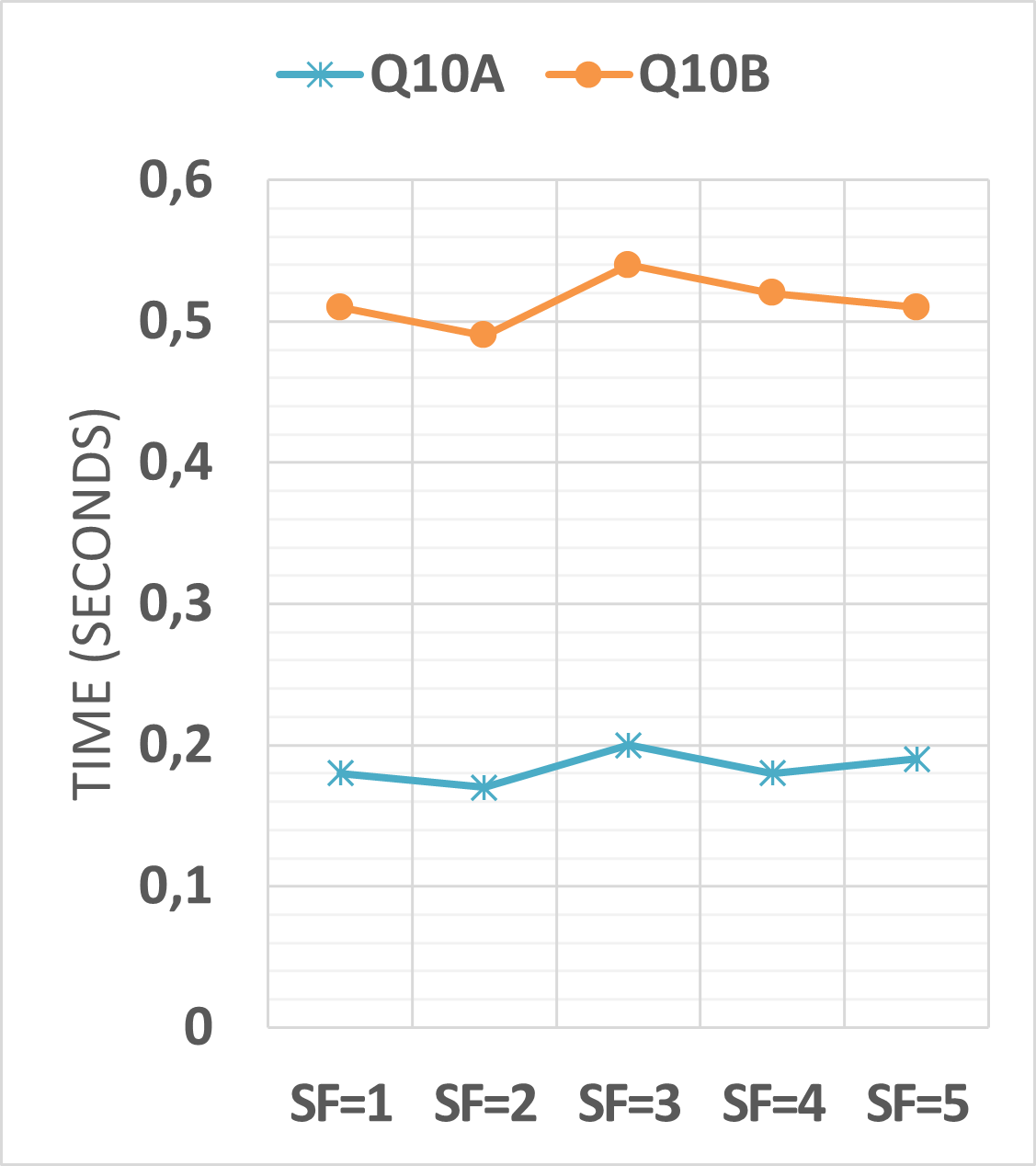} 
                \captionsetup{justification=centering}
                \captionof{figure}{Task \#10 \newline response times}
                \label{fig:Q10}
            \end{minipage}
        \end{figure*}

\begin{figure}[hbt]
    \centering
    \includegraphics[width=.7\textwidth]{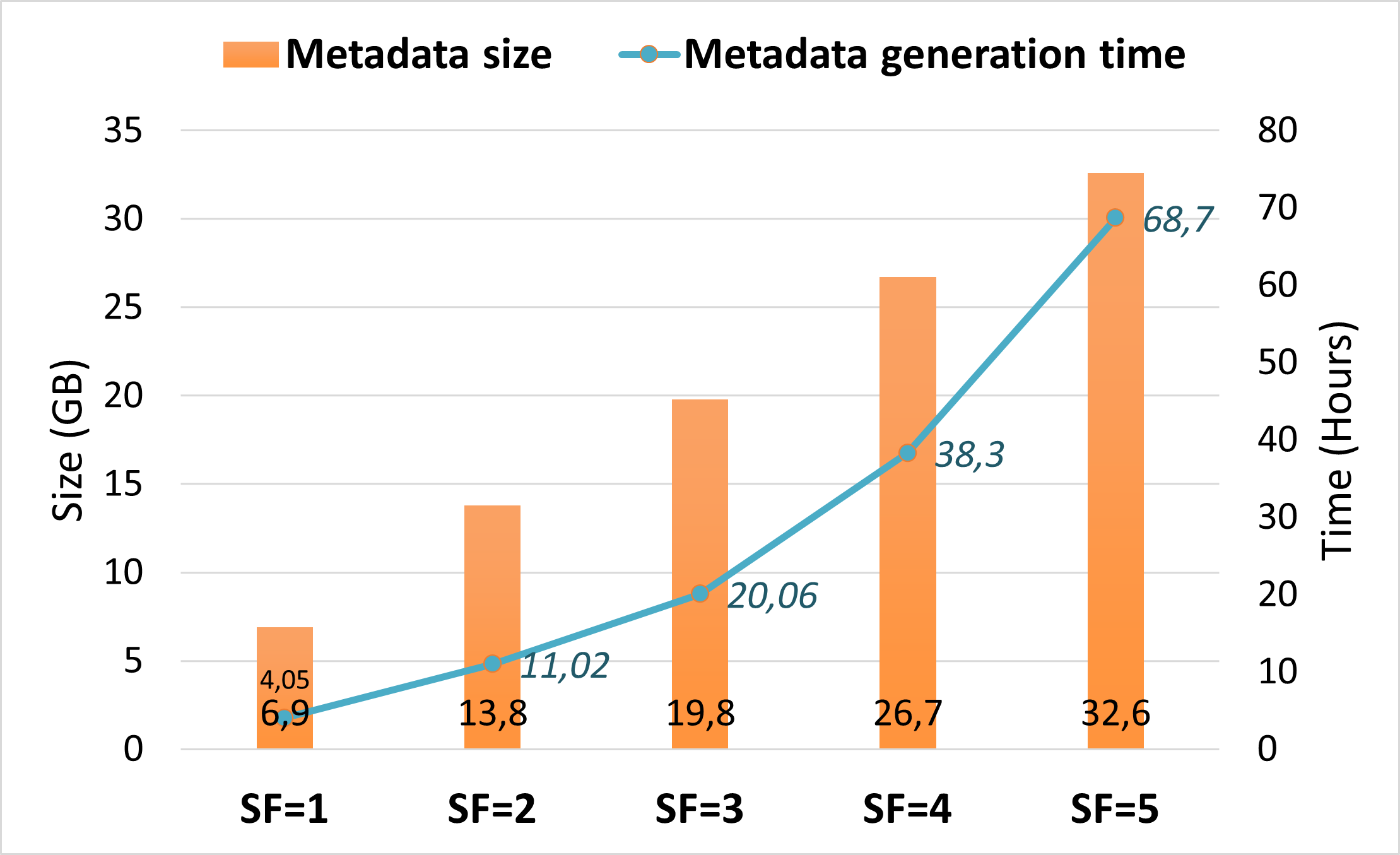}
    \caption{Metadata size and generation time}
    \label{fig:metadata.scale}
\end{figure}

The results achieved with \textsc{DLBench} show that AUDAL scales quite well (Figures~\ref{fig:Q1}-\ref{fig:Q10}). Almost all task response times are indeed either constant (Tasks \#3, \#7, \#8, \#9 and \#10) or grow linearly with $SF$ (Tasks \#1, \#2, and \#4 to \#6).  
In addition, we observe that except Task \#6 (that takes up to 92 seconds), all execution times are reasonable considering the modest capabilities of our hardware setup. 
Eventually,  metadata generation time and size scale linearly and almost-linearly, respectively (Figure~\ref{fig:metadata.scale}). We can also see that metadata amount to about half the volume of raw data, which illustrates how extensive AUDAL's metadata are. 
We observe some fluctuations in the results, with sometimes negative slopes while  $SF$ increases. Such variations are due to external, random factors such as network load or the Java garbage collector starting running. However, the influence on the runtime is negligible (of the order of a tenth of a second) and is only visible on simple queries that run in half a second. 

\section{Conclusion}
\label{sec:conclusion}

In this paper, we introduce \textsc{DLBench}, a benchmark for data lakes with textual and/or tabular contents. To  the best of our knowledge, 
\textsc{DLBench} is the first data lake benchmark. 
\textsc{DLBench} features: 1)~a data model made of a corpus of long, textual documents on one hand, and a set of raw tabular data on the other hand; 2)~a query model of twenty query instances across ten different tasks; 3)~three relevant metrics to assess and compare data lake implementations; and 4)~an execution protocol. 
Finally, we demonstrate the use of \textsc{DLBench} by assessing the AUDAL data lake~\cite{Scholly2021}, highlighting that the AUDAL system scales quite well, especially for data retrieval and tabular data querying.  

Future works include an extension of the structured part of \textsc{DLBench}'s data model with an alternative, larger dataset. Another enhancement of \textsc{DLBench} could 
consists in providing an overview of value distributions in generated data. Finally, we plan to perform a comparative study of existing data lake systems using \textsc{DLBench}.

\section*{Acknowledgments}
P.N. Sawadogo's PhD is funded by the Auvergne-Rhône-Alpes Region through the AURA-PMI project.

%
%
%
%

\end{document}